\documentstyle[12pt,html,epsf,psfig]{article}

\begin{document}
\title{MATTERS OF GRAVITY, The newsletter of the APS Topical Group on 
Gravitation}
\begin{center}
{ \Large {\bf MATTERS OF GRAVITY}}\\ 
\bigskip
\hrule
\medskip
{The newsletter of the Topical Group on Gravitation of the American Physical 
Society}\\
\medskip
{\bf Number 27 \hfill Spring 2006}
\end{center}
\begin{flushleft}

\tableofcontents
\vfill
\section*{\noindent  Editor\hfill}

Jorge Pullin\\
\smallskip
Department of Physics and Astronomy\\
Louisiana State University\\
Baton Rouge, LA 70803-4001\\
Phone/Fax: (225)578-0464\\
Internet: 
\htmladdnormallink{\protect {\tt{pullin-at-lsu.edu}}}
{mailto:pullin@lsu.edu}\\
WWW: \htmladdnormallink{\protect {\tt{http://www.phys.lsu.edu/faculty/pullin}}}
{http://www.phys.lsu.edu/faculty/pullin}\\
\hfill ISSN: 1527-3431
\begin{rawhtml}
<P>
<BR><HR><P>
\end{rawhtml}
%{\bf \Large Contents:}
\end{flushleft}
\pagebreak
\section*{Editorial}

As announced in the last edition, this is the last number I edit.
From now on editorship will fall upon our recently elected vice-chair,
David Garfinkle. We all wish David good luck with his new 
responsibilities.

In addition to the election of vice-chair, Alessandra Buonanno and Bob
Wagoner were elected to the executive committee. The two modifications
to the bylaws (creation of the membership coordination post and
correction of a typo in the number of members of the nominating
committee) were approved overwhelmingly in the ballot.

The next newsletter is due September 1st. All issues are available in
the WWW:\\\htmladdnormallink{\protect
  {\tt{http://www.phys.lsu.edu/mog}}} {http://www.phys.lsu.edu/mog}\\
The newsletter is available for Palm Pilots, Palm PC's and web-enabled
cell phones as an Avantgo channel. Check out
\htmladdnormallink{\protect {\tt{http://www.avantgo.com}}}
{http://www.avantgo.com} under technology$\rightarrow$science.  A
hardcopy of the newsletter is distributed free of charge to the
members of the APS Topical Group on Gravitation upon request (the
default distribution form is via the web) to the secretary of the
Topical Group.  It is considered a lack of etiquette to ask me to mail
you hard copies of the newsletter unless you have exhausted all your
resources to get your copy otherwise.

If you think a topic should be covered by the newsletter you are
strongly encouraged to contact the relevant correspondent.  If you
have comments/questions/complaints about the newsletter email me. Have
fun.

\hfill Jorge Pullin
%\vfill\eject

\bigbreak

\vspace{-0.8cm}
\parskip=0pt
\section*{Correspondents of Matters of Gravity}
\begin{itemize}
\setlength{\itemsep}{-5pt}
\setlength{\parsep}{0pt}
\item John Friedman and Kip Thorne: Relativistic Astrophysics,
\item Bei-Lok Hu: Quantum Cosmology and Related Topics
\item Gary Horowitz: Interface with Mathematical High Energy Physics and
String Theory
\item Beverly Berger: News from NSF
\item Richard Matzner: Numerical Relativity
\item Abhay Ashtekar and Ted Newman: Mathematical Relativity
\item Bernie Schutz: News From Europe
\item Lee Smolin: Quantum Gravity
\item Cliff Will: Confrontation of Theory with Experiment
\item Peter Bender: Space Experiments
\item Jens Gundlach: Laboratory Experiments
\item Warren Johnson: Resonant Mass Gravitational Wave Detectors
\item David Shoemaker: LIGO Project
\item Peter Saulson: former editor, correspondent at large.
\end{itemize}
\section*{Topical Group in Gravitation (GGR) Authorities}
Chair: Jorge Pullin; Chair-Elect: \'{E}anna Flanagan; Vice-Chair: 
Dieter Brill. 
Secretary-Treasurer: Vern Sandberg; Past Chair: Jim Isenberg;
Delegates:
Bei-Lok Hu, Sean Carroll,
Bernd Bruegmann, Don Marolf, 
Vicky Kalogera, Steve Penn.
\parskip=10pt

\vfill
\eject

\section*{\centerline {
GGR program at the APS April meeting in Dallas}}
\addtocontents{toc}{\protect\medskip}
\addtocontents{toc}{\bf GGR News:}
\addcontentsline{toc}{subsubsection}{\it
GGR program at the APS April meeting in Dallas}
\begin{center}
Jorge Pullin, Louisiana State University
\htmladdnormallink{pullin-at-lsu.edu}
{mailto:pullin@lsu.edu}
\end{center}

We have an exciting GGR related program at the upcoming 
APS April meeting in Dallas, Texas, April 22-25 2006. 
Early registration deadline is February 24. 
Our chair elect, \'{E}anna Flanagan did a remarkable 
job putting this program together.

0. Plenary talk on LIGO Speaker: Gabriela Gonz\'alez 

I. Ground-based Gravitational Wave Detection (Saturday, April 22, 1:30pm)\\
Chair: Benjamin Owen (joint with Topical Group on Precision Measurements)\\ 
Mike Zucker --- Status of LIGO \\
Patrick Brady --- Results from LIGO observations I \\
Patrick Sutton --- Results from LIGO observations II 

II. Experimental Tests of General Relativity (Saturday, April 22, 3:30pm)\\
Chair: Marc Favata \\
John Anderson --- Anomalous Acceleration of Pioneer 10 and 11 \\
Slava Turyshev --- Science, Technology and Mission Design for the 
Laser Astrometric Test of Relativity \\
Eric Adelberger --- Tests of the gravitational inverse-square law 
at the dark-energy length scale 

III. Theories of Gravity, Dark Energy and Cosmology 
(Sunday, April 23, 10:30am)\\
Chair: Sean Carroll (joint with Division of Particles and Fields) \\
Shamit Kachru --- String Theory and Cosmology \\
Nima Arkani-Hamed --- Implications of the Accelerating Universe 
for Fundamental Physics \\
Roman Scoccimarro --- Differentiating between Modified Gravity 
and Dark Energy \\

IV. Precision Cosmology (Sunday, April 23, 1:15pm)\\
Chair: John Beacom (joint with Division of Astrophysics) \\
Lyman Page --- Recent Results from WMAP \\
Josh Frieman --- Probing Dark Energy with Galaxy Clusters \\
Daniel Eisenstein --- Acoustic Oscillations in Galaxy Large-Scale Structure \\

V. Advances in Numerical Relativity (Sunday, April 23, 3:15pm)\\
Chair: Deirdre Shoemaker (joint with Division of Computational Physics) \\
Frans Pretorius --- Simulations of Binary Black Hole Mergers \\
Larry Kidder --- Numerical Simulation of Binary Black Holes \\
Thomas Baumgarte --- Neutron Stars in Compact Binaries \\
David Garfinkle --- Numerical simulations of generic singularities \\

VI. Gravitational Wave Sources and Phenomenology (Monday, April 24, 10:45am)\\
Chair: Gabriela Gonz\'alez (joint with Division of Astrophysics) \\
Curt Cutler --- Overview of LISA Science \\
Alessandra Buonanno --- Source-modeling, detection and science of 
gravitational waves from compact binaries \\
Coleman Miller --- Gravitational Radiation from 
Intermediate-Mass Black Holes 

VII. Heineman prize session (Tuesday, April 25, 10:45am)\\
Chair: Pierre Sikivie (joint with Division of Particles and Fields)\\
Citation: "For constructing supergravity, the first supersymmetric 
extension of Einstein's theory of general relativity, and for 
their central role in its subsequent development."\\
Sergio Ferrara --- Current topics in the theory of supergravity \\
Daniel Freedman --- Supergravity and the AdS/CFT Correspondence \\
P. van Nieuwenhuizen --- Supergravity \\

VIII. Focus session on Numerical Relativity\\
Lead Speaker: Greg Cook --- Status of Initial Data for Binary Black Hole Collisions 

IX. Focus session on Space-Based Gravitational Wave Detection\\
Lead Speaker: Neil Cornish --- The LISA Observatory: Preparing 
for a bountiful harvest 

X. Focus session on Recent Results in Quantum Gravity\\
Lead Speaker: Lee Smolin --- Physical predictions from loop quantum gravity

\section*{\centerline {
We hear that...}}
\addcontentsline{toc}{subsubsection}{\it
We hear that..., 100 years ago, by Jorge Pullin}
\begin{center}
Jorge Pullin, LSU
\htmladdnormallink{pullin-at-lsu.edu}
{mailto:pullin@lsu.edu}
\end{center}

Bruce Allen, Peter Fritschel and Don Marolf were elected fellows of the APS.

Ruth Gregory won the Maxwell Medal of the Institute of Physics (UK).

Alessandra Buonanno and Nergis Mavalvala were awarded  Sloan fellowships.

Hearty Congratulations!

\section*{\centerline {
100 Years ago}}
\parskip=3pt
\begin{center}
Jorge Pullin
\htmladdnormallink{pullin-at-lsu.edu}
{mailto:pullin@lsu.edu}
\end{center}
An English version of 
``On the dynamics of the electron'' by Henri Poincar\'e, is available at
 
\htmladdnormallink{http://www.phys.lsu.edu/mog/100}
{http://www.phys.lsu.edu/mog/100}
\vfill
\eject

\section*{\centerline {
What's new in LIGO}}
\addtocontents{toc}{\protect\medskip}
\addtocontents{toc}{\bf Research Briefs:}
\addcontentsline{toc}{subsubsection}{\it  
What's new in LIGO, by David Shoemaker}
\begin{center}
David Shoemaker, LIGO-MIT
\htmladdnormallink{dhs-at-ligo.mit.edu}
{mailto:dhs@ligo.mit.edu}
\end{center}

In an important sense, LIGO has recently turned a corner in its
history: It has moved from the commissioning to the observation
phase.

Since the last MOG, a number of technical issues have been addressed
in all three interferometers. Increases in the laser input power,
tuning of the system which compensates for the thermally-induced
focusing in optics, work on reducing scattered light paths and
acoustic excitation of optic motion, and control-law optimizations
are among the specific efforts. This has both improved the strain
sensitivity as well as increased the duty cycle of operation of the
instruments.

The result is that the two 4km interferometers exceed the
performance promised in the 1995 LIGO Science Requirements Document
of a sensitivity of $10^{-21}$ in strain for a 100 Hz bandwidth,
with the 2km interferometer also functioning well given its shorter
length. The LIGO Scientific Collaboration had given its agreement to
proceeding with the definitive S5 science run at the August LSC
meeting, and the NSF Annual Review of LIGO held in November 2005
also confirmed that the target sensitivity was achieved.

The S5 science run, underway since mid-November 2005, is intended to
collect one year of integrated coincidence data between the two LIGO
Observatories. We plan to take breaks in observation from time to
time to implement small improvements, and repair any equipment that
breaks down during the run. Some observation time is lost to
maintenance, and the first stage of construction of an Outreach
center at Livingston will impact the day-time duty cycle of that
instrument for the beginning of the run. All factors taken into
account, we plan to run for about 1.5 years to accumulate these
data.

Online (near real time) data analysis tools are characterizing the
data on-the-fly, helping the staff optimize the instruments and
recognizing quickly any problems that need to be addressed. The four
basic searches, for signals with the character of bursts, a
stochastic background, periodic or quasi-periodic, and binary
inspiral signatures, are being applied to the data, and the LSC
plans to keep the analysis process active continuously throughout
the run.

Analysis continues on the previous science runs, with better upper
limits on a variety of sources established and new techniques
exercised which will be employed also in the S5 analysis. Papers
have appeared or accepted on searches for periodic sources (``First
all-sky upper limits from LIGO on the strength of periodic
gravitational waves using the hough transform'') and on burst
searches `triggered' by GRB signals (``Search for gravitational
waves associated with the gamma ray burst GRB030329 using the LIGO
detectors''). A variety of other publications is in preparation;
searching on gr-qc for `the LIGO Scientific Collaboration' is an
effective way to stay up-to-date.

Advanced LIGO has also made strides forward. The characterization of
the mirror suspensions and of the seismic isolation systems has
progressed, and full-scale prototypes of the suspensions and seismic
isolation will converge for integrated testing at the MIT LASTI
facility in the coming months. The 40m interferometer test bed at
Caltech has successfully demonstrated the length control scheme for
the Advanced LIGO signal- and power-recycled Fabry-Perot Michelson
configuration. Extensive modeling has helped our understanding of
thermal compensation, possible parametric excitation of mirror
modes, and the requirements to be placed on the mirror figure. Four
of the actual to-be-installed fused silica test masses, contributed
by the UK, have been delivered and will go through a pathfinding
process to identify polishing and coating techniques.

Advanced LIGO has also appeared in the recent 2007 budget materials
from the OMB and the NSF as indicated for an FY2008 start. Although
the official decision is still in the future, this is a strong
indication of the support from the NSF and the interest in the
government to support this field, and an affirmation of the LSC's
very successful effort to advance the astrophysics and the
instrument science to the point where all agree that this is timely.
A baseline review will be held in late May 2006 to confirm the cost,
schedule, risk handling, and technical plans, and we hope to be very
busy with preparing for the start of the project from that point
onward.

Last but not least: In the last MOG, we mentioned that the LIGO
Laboratory was involved in a search for a new director. Jay Marx,
formerly of LBNL, has accepted the position of Director, and we
welcome him warmly to the Lab and the field.

\vfill
\eject
\section*{\centerline {LISA Pathfinder}}
\addtocontents{toc}{\protect\smallskip}
\addcontentsline{toc}{subsubsection}{\it
LISA Pathfinder, by Paul McNamara}
\begin{center}
Paul McNamara, ESTEC-ESA
\htmladdnormallink{Paul.McNamara-at-esa.int}
{mailto:Paul.McNamara@esa.int}
\end{center}
\parindent=0pt
\parskip=5pt

LISA Pathfinder (formerly known as SMART-2), the second of the ESA
Small Missions for Advanced Research in Technology, is a dedicated
technology demonstrator for the joint ESA/NASA Laser Interferometer
Space Antenna (LISA) mission.

The technologies required for LISA are many and extremely challenging.
This coupled with the fact that some flight hardware cannot be tested
on ground due to the earth induced noise, led to the LISA Pathfinder
(LPF) mission being implemented to test the critical LISA technologies
in a flight environment. The scientific objective of the LISA
Pathfinder mission consists then of the first in-flight test of
gravitational wave detection metrology.

LISA Pathfinder carries two payloads, the European provided LISA
Technology Package (LTP) and the NASA provided Disturbance Reduction
System - Precision Flight Control Validation (DRS PCFV), formerly
known as the DRS.

{\bf Mission Goals}

The mission goals of the LTP can be summarized as:

$\bullet$ demonstrating that a test-mass can be put in pure
gravitational free-fall within one order of magnitude of the
requirement for LISA. The one order of magnitude rule applies also to
frequency, thus the flight test of the LTP on LPF is considered
satisfactory if free-fall of one TM is demonstrated to within
\begin{equation}
S_a^{1/2}(f) \leq 3\times 10^{-14}\left[ 1 
+\left({f \over 3 {\rm mHz}}\right)^2\right] {\rm ms}^{-2} /\sqrt{\rm Hz}
\end{equation}
over the frequency range, $f$, of 1 to 30 mHz.

$\bullet$ demonstrating laser interferometry with a free-falling
mirror (test mass of LTP) with displacement sensitivity meeting the
LISA requirements over the LTP measurement bandwidth. Thus the flight
test of LTP is considered satisfactory if the laser metrology
resolution is demonstrated to within
\begin{equation}
S_{\delta x} ^{1/2}(f) = 9.1\times 10^{-12}\left[ 1 
+\left({f \over 3 {\rm mHz}}\right)^{-2}\right] {\rm m} /\sqrt{\rm Hz}
\end{equation}
over the frequency range, $f$, of 1 to 30 mHz.

$\bullet$ assessing the lifetime and reliability of the micro-Newton
thrusters, lasers and optics in a space environment.

{\bf LTP}

The basic idea behind the LTP is that of squeezing one arm of LISA
from $5\times 10^6$ km to a few centimeters and placing it on board a
single S/C. Thereby the key elements are two nominally free flying
test masses (TM), and a laser interferometer whose purpose is to read
the distance between the TM's (Figure 1).  

\begin{figure} 
\centerline{\psfig{file=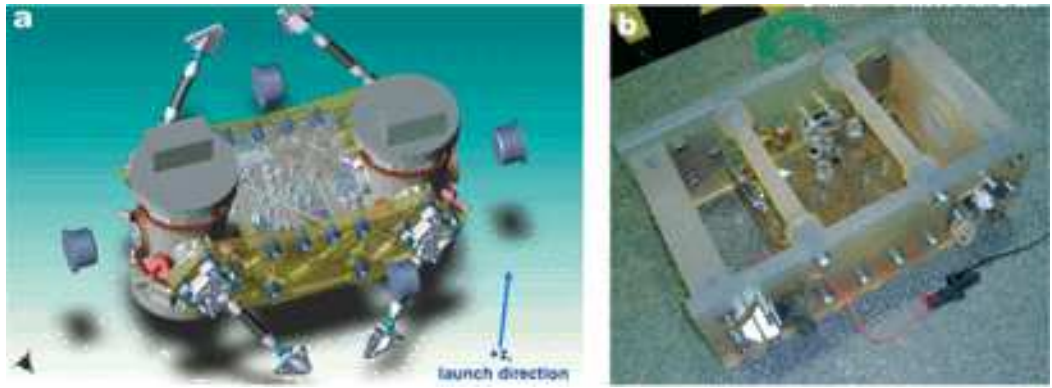,height=2in}}
\caption{
  a) CAD drawing of the LISA Technology Package showing the two vacuum
  enclosures housing the test masses, and the optical bench
  interferometer (OBI), b) photograph of the EM of the OBI (vacuum
  enclosures replaced with {\em end plates}).}
\end{figure}

The two tests masses are surrounded by their position sensing
electrodes. This position sensing provides the information to a
"drag-free" control loop that, via a series of micro-Newton thrusters,
keeps the spacecraft centered with respect to some fiducial point.

In LISA, as in LPF, each spacecraft hosts two test-masses. However
these two test-masses belong to different interferometer arms. This
has an important consequence for the logic of the spacecraft control.
The baseline defined by the system level study for LISA, sees a
control logic where the spacecraft is simultaneously centered on both
test-masses. However the spacecraft follows each test-mass only along
the axis defined by the incoming laser beam. The remaining axes have
to be controlled by a capacitive suspension (or by some other
controlled actuation scheme). On LPF however, in order to be able to
measure differential acceleration, the sensitive axes of the two
test-masses have to be aligned. This forces one to develop a
capacitive suspension scheme that carries one or both test-masses
along with the spacecraft, including along the measurement axis, while
still not spoiling the meaningfulness of the test.

In LISA, the proper distance between the two free-falling test masses
at the end of the interferometer arms is measured via a three step
process; by measuring the distance between one test mass and the
optics bench (known as the {\em local measurement}), by measuring the
distance between optics benches (separated by 5 million kilometers),
and finally be measuring the distance between the other test mass and
its optics bench. In LISA Pathfinder, the optical metrology system
essentially makes two measurements; the separation of the test masses,
and the position of one test mass with respect to the optics bench.
The latter measurement is identical to the LISA local measurement
interferometer, thereby providing an in-flight demonstration of
precision laser metrology directly applicable to LISA.

In LISA and in LPF, charging by cosmic rays is a major source of
disturbance, thereby each test-mass carries a non contacting charge
measurement and neutralization system based on UV photoelectron
extraction. An in-flight test of this device is then obviously a key
element of the overall LPF test.

{\bf Disturbance Reduction System - Precision Control Flight Validation}

The DRS-PCFV is a NASA provided payload to be flown on the LISA
Pathfinder spacecraft. When first proposed, the DRS payload closely
resembled the LTP, namely in that it consisted of two inertial sensors
with associated interferometric readout, as well as the drag-free
control laws and micro-Newton colloidal thrusters, although the
technologies employed were different from the LTP. However, due to
budgetary constraints, the DRS was de-scoped, and now consists of the
micro-Newton colloidal thrusters, drag-free and attitude control
system (DFACS), and a micro-processor. The DRS-PCFV will now use the
LTP inertial sensors as its drag-free sensors.

The primary goal of the DRS-PCFV is to maintain the position of the
spacecraft with respect to the proof mass to within 10nm$/\sqrt{\rm
  Hz}$ over the frequency range of 1-30mHz.

{\bf Launch and orbit}

LISA Pathfinder is due to be launched in October 2009 on-board a
dedicated launcher. Rockot is presently the baseline vehicle, while
VEGA is considered the target vehicle that will be used if available.
The spacecraft and propulsion module (Figure 2) are injected into a
low earth orbit (200 x 900km), from which, after a series of apogee
raising burns, will enter a transfer orbit towards the first Sun-Earth
Lagrange point (L1). After separation from the propulsion module, the
LISA Pathfinder spacecraft will be stabilized using the micro-Newton
thrusters, entering a Lissajous orbit around L1 (500,000km by
800,000km orbit).

\begin{figure}
\centerline{\psfig{file=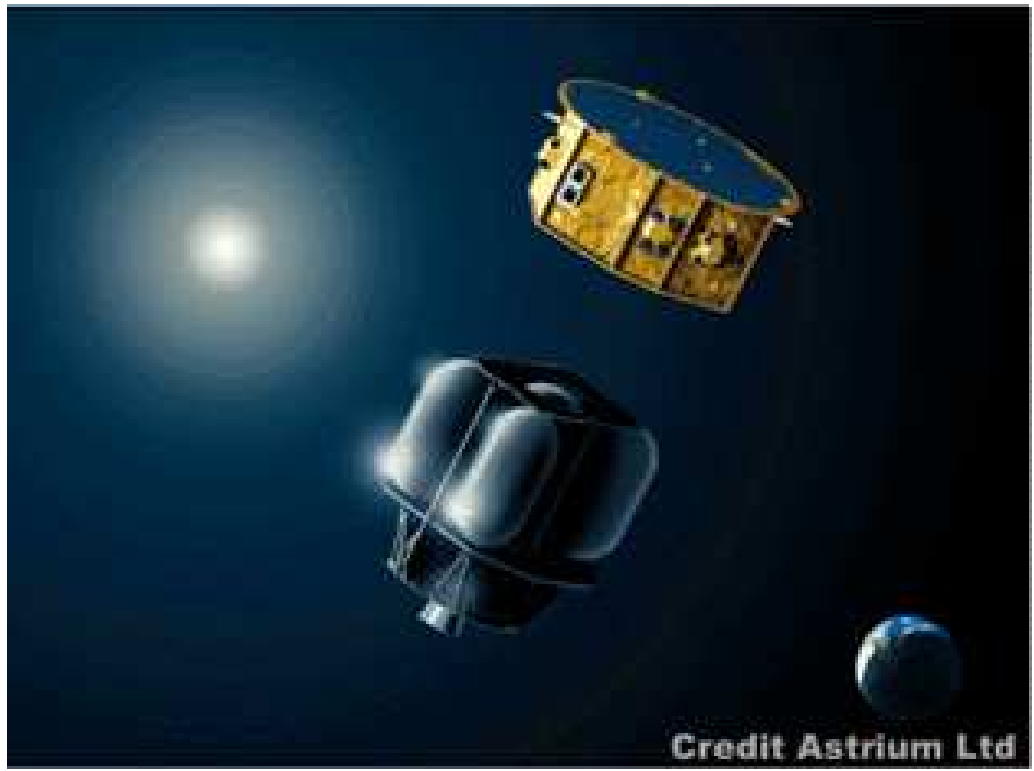,height=4in}}
\caption{The LISA Pathfinder spacecraft separating from its propulsion
module.}
\end{figure}

Following the initial on-orbit check-out and instrument calibration,
the in-flight demonstration of the LISA technology will take place in
the first half of 2010. The nominal lifetime of the mission is 180
days, this includes the LTP operations, the DRS operations, and a
period of joint operations when the LTP will control the DRS
thrusters.

{\bf Status}

LISA Pathfinder is currently in Implementation Phase. The contract
with the prime industrial contractor, Astrium UK, was signed in May
2004. During the last year, all ITTs for spacecraft subcontractors
have been issued.

In October 2004, the Science Program Council (SPC) approved the LTP
Multi-Lateral Agreement, detailing the national agency
responsibilities for the construction of the LTP. All subcontracts for
the LTP have started.

The project has also successfully completed a number of significant
agency level reviews over the last year, including the Technology
Readiness Review, the LTP Preliminary Design Review, System
Preliminary Design Review, and the Mission Preliminary Design Review.
Also, all the LTP subsystems have undergone PDR within the last year.

With the deletion of the GRS from the DRS, it was recommended that the
DRS undergo a joint ESA/NASA delta-Critical Design Review ($\delta$-CDR)/Risk
Review. This was completed successfully in January 2006.

The first LTP subsystem flight hardware is due to be delivered to the
LTP Architect (Astrium GmbH) during the third quarter 2006. The
delivery of the assembled and tested LTP instrument to the prime
contractor is scheduled for July 2008.

\vfill\eject

\section*{\centerline {Recent progress in binary black hole simulations}}
\addtocontents{toc}{\protect\smallskip}
\addcontentsline{toc}{subsubsection}{\it
Recent progress in binary black hole simulations, by Thomas Baumgarte}
\begin{center}
    Thomas Baumgarte, Bowdoin College \\
\htmladdnormallink{tbaumgar-at-bowdoin.edu}
{mailto:tbaumgar@bowdoin.edu}
\end{center}
\parindent=0pt
\parskip=5pt

\def\alt{
\mathrel{\raise.3ex\hbox{$<$}\mkern-14mu\lower0.6ex\hbox{$\sim$}}
}

\def\etal{{\it et.al.~}}

\bigskip

The past year has seen dramatic progress in numerical relativity
simulations of binary black holes.  A number of groups have reported
significant advances and are now able to model the binary inspiral,
coalescence and merger together with the emitted gravitational wave
signal.

Simulating binary black holes has been a long-standing problem because
it poses a number of ``grand challenges''.  An incomplete list of
these challenges includes the following
\begin{itemize}
\item Einstein's equations form a complicated, coupled set of 
non-linear PDEs, and it is far from clear which form of these
equations is most suitable for numerical implementation.
\item Somewhat related is the coordinate freedom, and the question
what coordinate (or gauge) conditions lead to a non-pathological 
evolution.
\item Black holes contain singularities, which can have very unfortunate
consequences for numerical simulations.
\item The individual black holes are much smaller than the wavelength
of the emitted gravitational radiation.  The resulting range in
length-scales is difficult to accommodate in numerical simulations.
\end{itemize}
The different groups have approached these issues in different ways.

Pretorius (2005) first announced his new results at a numerical
relativity workshop at the Banff International Research Station.
Departing from numerical relativity convention he does not integrate a
``3+1'' decomposition of Einstein's equations that separates spatial and
timelike parts, but instead discretizes the four-dimensional spacetime
equations and their second derivatives directly.  As suggested by a
number of previous authors he introduces gauge source functions
$H_{\mu}$, in terms of which Einstein's equations reduce to wave
equations for the components of the spacetime metric.  In this
formalism the coordinates are fixed through the gauge source functions
(instead of the lapse and shift in 3+1 formalisms).  Pretorius chooses
$H_t$ to satisfy a somewhat ad-hoc wave equation and $H_i = 0$ (which
is related to spatial harmonic coordinates $H^i = 0$).

Pretorius uses black hole excision, whereby the black hole interior is
removed from the computational grid.  This is justified since the
event horizon disconnects the interior causally from from the
exterior.  He also uses adaptive mesh refinement (AMR), which
automatically allocates additional gridpoints in regions where they
are needed to resolve small scale structures.

Several other features of his code are worth pointing out.  The
spatial coordinates are compactified, so that physically correct outer
boundaries can be imposed at spatial infinity.  He also introduces
some numerical dissipation to control high-frequency instabilities,
and added some ``constraint-damping'' terms that proved crucial for
simulations of black holes.

With this code Pretorius has been able to simulate -- without
encountering a numerical instability -- the inspiral and coalescence
of black hole binaries through merger until late stages of the
ring-down as the remnant settles into equilibrium.  This is remarkable
progress indeed.  Figure 1 shows the trajectory of an inspiraling
black hole binary and a gravitational waveform from his very recent
calculation that adopts the state-of-the-art initial data of Cook and
Pfeiffer (2004).

Following Pretorius' success four other groups (Campanelli \etal
(2005), Baker \etal (2005), Diener \etal (2005) and most recently
Herrmann \etal (2006)) have announced significant progress in their
binary black hole simulations.  All of these calculations have several
features in common.  They all use finite-difference implementations of
the BSSN equations\footnote{In the BSSN formalism a set of auxiliary
  connection functions $\Gamma^i$ is introduced that simplify the
  three-dimensional Ricci tensor ${}^{(3)}R_{ij}$ in the same way as
  the $H^{\mu}$ simplify the four-dimensional Ricci tensor
  ${}^{(4)}R_{\mu\nu}$ in the formalism adopted by Pretorius.}, which
are based on a 3+1 formalism in contrast to Pretorius'
four-dimensional approach.  They also use very similar gauge
conditions, namely ``1 + log'' slicing for the lapse, and a ``driver''
implementation of ``Gamma-freezing''.  Finally they all use
``puncture'' initial data, which are constructed by absorbing the
singular terms in the black hole interior into an analytical
expression and solving for regular corrections.

\begin{figure}[t]
\centerline{
\psfig{file=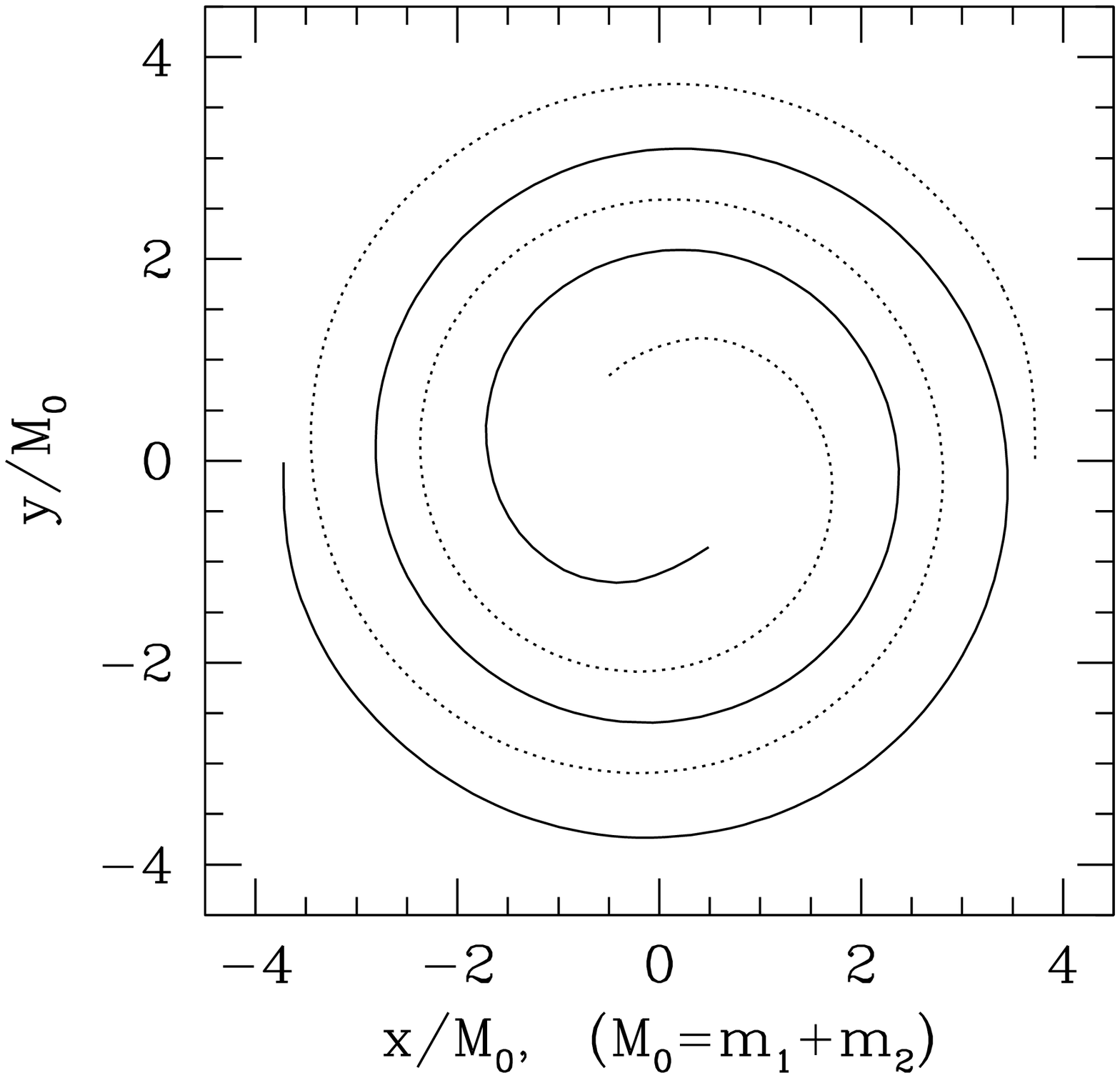,width=3in}~~~
\psfig{file=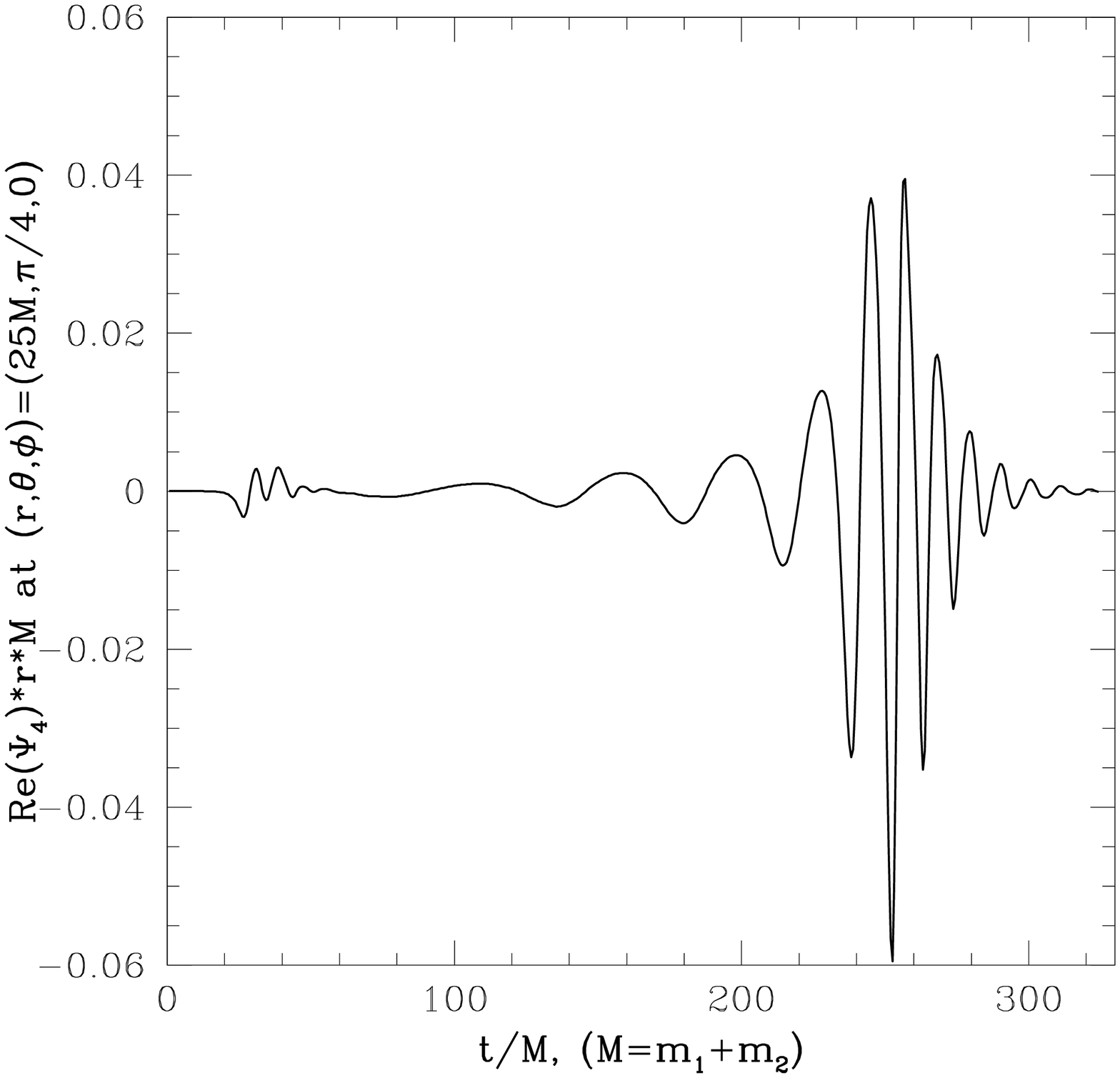,width=2.7in}}
\caption{The left panel shows black hole trajectories in a recent
binary black hole simulation of Pretorius (2006, private
communication), starting with an initial data configuration of Cook
and Pfeiffer (2004).  The right panel shows the corresponding
``gravitational waveform'' $Re(\psi_4)$.}
\end{figure}

The approach of Campanelli \etal (2005) and Baker \etal (2005) differs
from the others in that they do not excise the black hole interiors,
and instead continue to use the ``puncture'' approach to handle the
singularities during the evolution.  Campanelli \etal (2005) introduce
a new variable that is the inverse of the diverging term.  This new
term vanishes at the ``punctures'', and given suitable gauge
conditions all equations remain regular.  Baker \etal (2005) finite
difference the diverging term directly, but arrange the computational
grid in such a way that the singularity never encounters a gridpoint.
In situations with equatorial symmetry, when both singularities reside
on the equatorial plane, this can always be achieved simply by using
cell-centered differencing schemes.  The two calculations also use
different grid structures and differencing; Campanelli \etal (2005)
adopt 4th order differencing on a uniform grid but introduce a
``fish-eye'' coordinate that provides additional resolution for the
black holes, while Baker \etal (2005) use 2nd order differencing and
FMR in an inertial coordinate system.  Figure 2 shows a gravitational
wave form from Campanelli \etal, and a demonstration of energy
conservation from Baker \etal.  Both groups also report satisfactory
agreement with earlier ``Lazarus'' results which combines numerical 
relativity with perturbative techniques (e.g.~Baker \etal (2001)).

\begin{figure}[t]
\centerline{\psfig{file=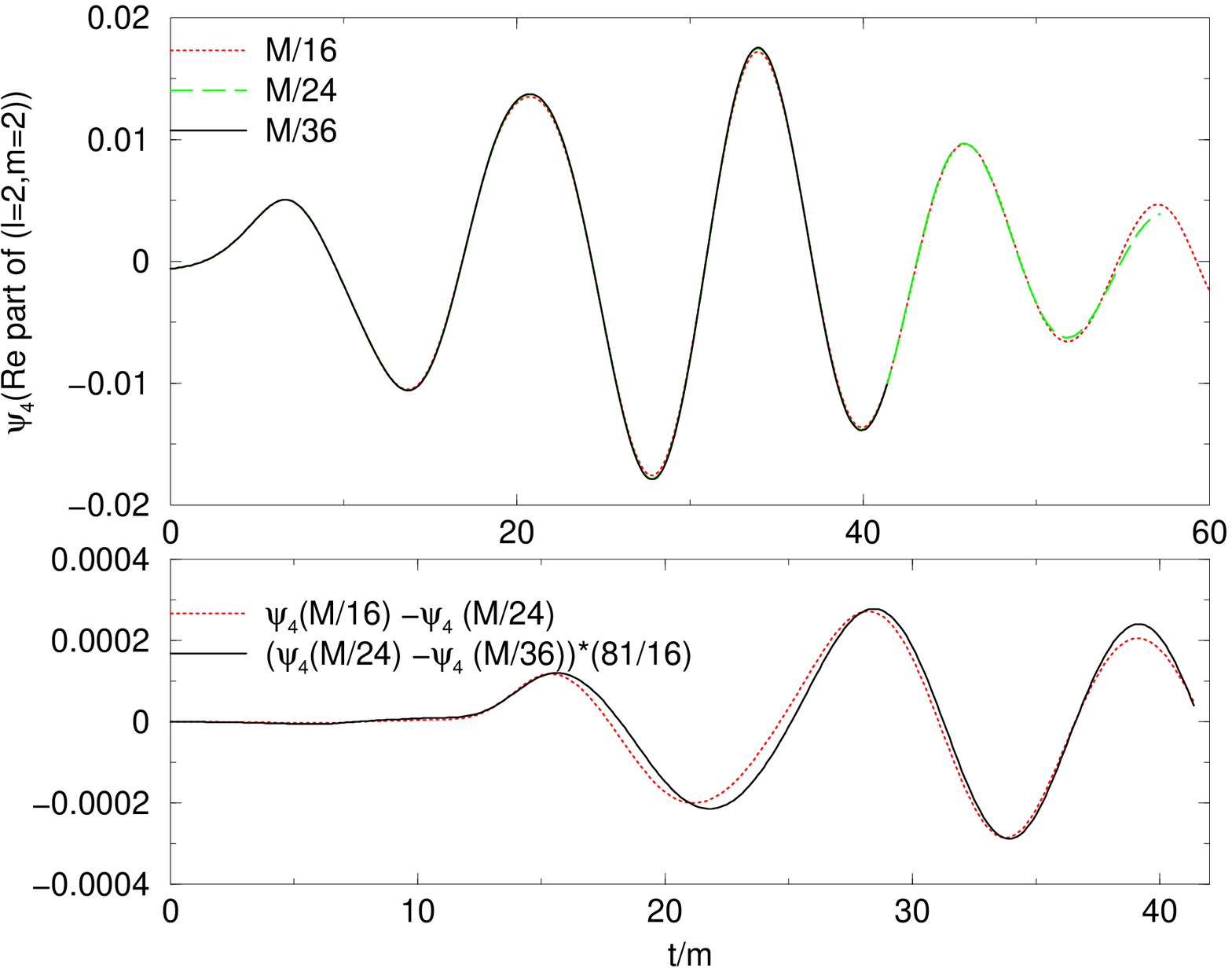,width=2.5in}\,
\psfig{file=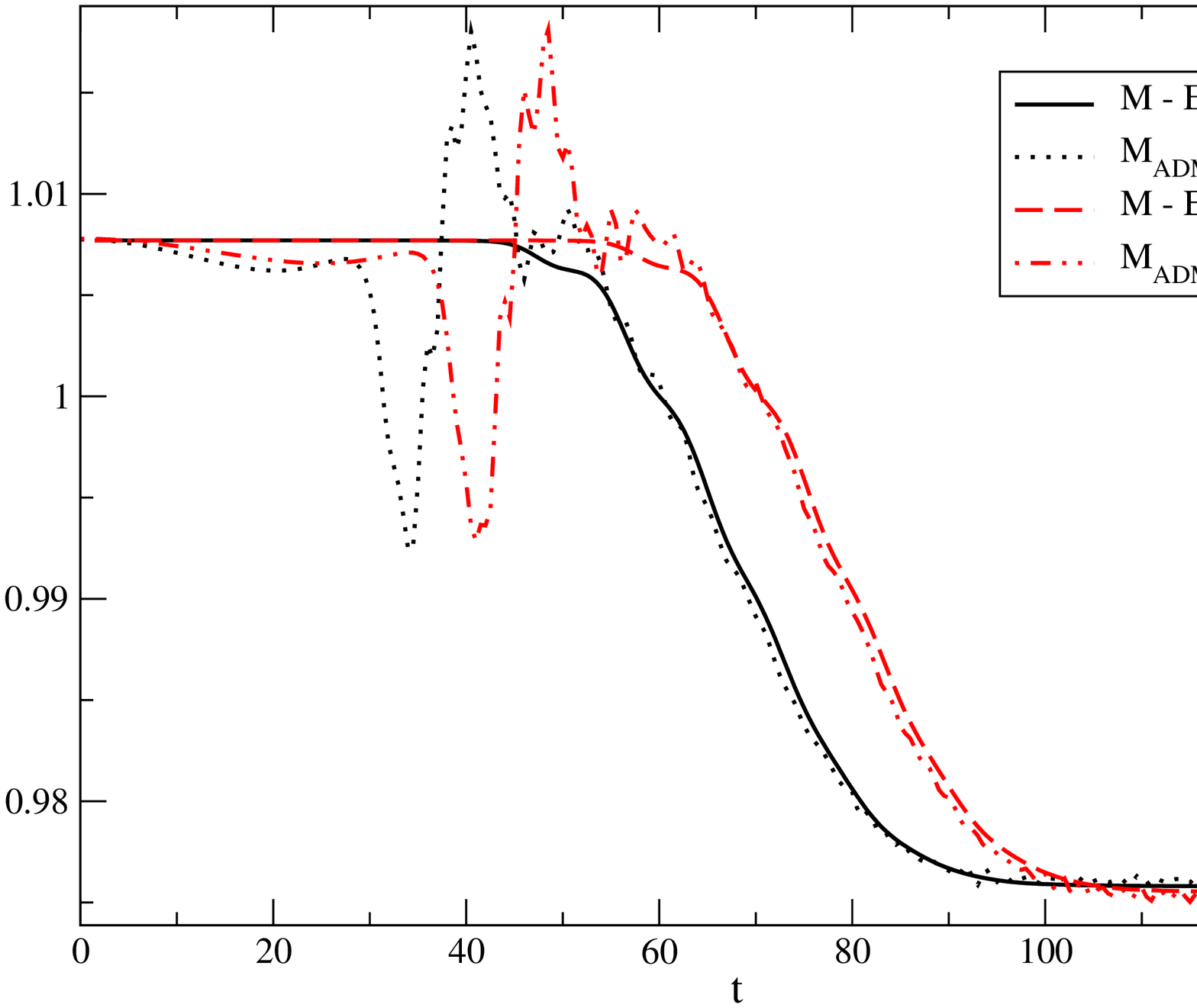,width=2.0in}}
\caption{Left Panel: The gravitational waveform $Re(\psi_4)$ in
the calculation of Campanelli \etal (2005).  This convergence test
demonstrates 4th order convergence.  Right Panel:  Demonstration
of energy conservation in the calculation of Baker \etal (2005).  The
initial energy $M$ minus the energy $E$ lost in gravitational radiation
agrees with the current total energy $M_{\rm ADM}$ to high accuracy.}
\end{figure}

Diener \etal (2005) report on impressive improvements of their earlier
results (Alcubierre \etal (2005); compare Br\"ugmann \etal (2004)).
Like Pretorius, they use black hole excision to eliminate the
curvature singularities in the black hole interior.  They use a fixed
mesh refinement (FMR) in a corotating coordinate system to resolve the
black holes.  A schematic of their black hole trajectories is shown in
the left panel of Figure 3.  Starting from an initial proper
separation of about 9 M the black holes spiral toward each other until
a common apparent horizon forms after about 1.5 orbits.  Diener \etal
(2005) also demonstrate that the spurious effect of finite difference
error on the binary orbit depends on the gauge condition.  All gauge
choices converge to the same physical solution, as expected, but at
finite resolution different choices may lead to either a widening or
closing of the orbit, which helps to explain earlier discrepancies
(compare Br\"ugmann \etal (2004), Alcubierre \etal (2005)).

The simulations of Campanelli \etal (2005) and Baker \etal (2005)
demonstrate that standard numerical relativity codes can handle binary
black holes with only very minor modifications, potentially opening
the field to a number of other groups.  Herrmann \etal (2006), for
example, adopt a technique very similar to that of Baker \etal (2005).
While all other simulations focus on equal-mass black holes they
consider unequal-mass black holes with mass ratios $q = M_1/M_2$
ranging from unity to 0.54.  Their calculation represents a first step
toward analyzing the effect of binary parameters -- including mass
ratios and black hole spins -- on the gravitational waveforms in fully
dynamical simulations (even if the astrophysical relevance of their
initial data is somewhat limited).  They also see evidence for
gravitational radiation recoil leading to a remnant ``kick'' (compare
the right panel of Figure 3).

\begin{figure}[t]
\centerline{
\psfig{file=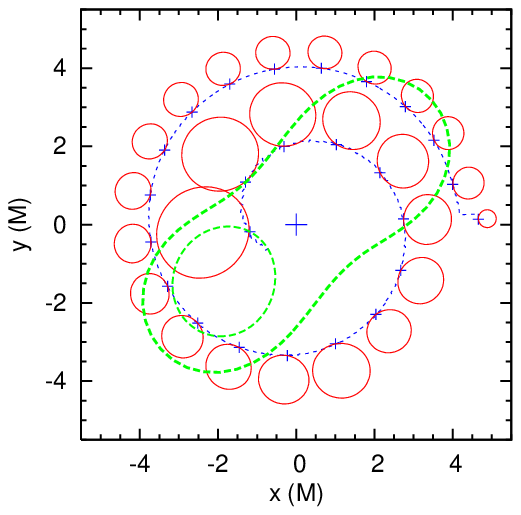,width=3.8in}\hspace*{-1in}
\psfig{file=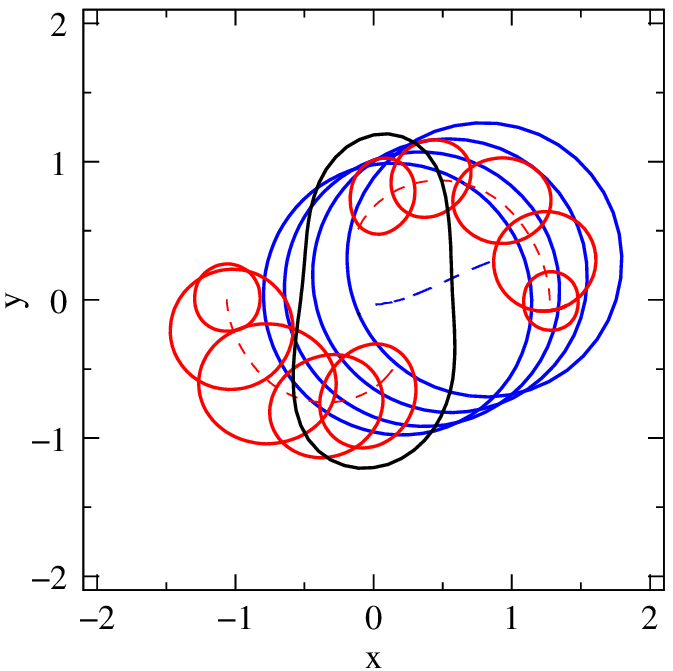,width=2.5in}}
\caption{ Left Panel: Motion of one of the black holes with time in
the simulation of Diener \etal (2005).  Cross-sections of the apparent
horizon (AH) with the equatorial plane are shown at intervals of
$\Delta t = 5 M$.  The apparent growth of the AHs with time is a pure
coordinate effect.  The first appearance of a common AH at $t = 124
M$, and the corresponding final detached AH, are shown as dashed
lines.  Right Panel: Snapshots of the apparent horizon locations for
the $q=0.85$ unequal-mass binary calculation of Herrmann \etal (2006).
The snapshots are taken every 4 $M_{\rm ADM}$ prior to merger (red)
and every 17 $M_{\rm ADM}$ after merger (blue).  The trajectories of
the common horizons' centers are shown as a dashed lines.}
\end{figure}

All of these calculations can clearly be improved in multiple ways.
However, especially comparing with the situation just a year ago, it
is quite remarkable and reassuring that different groups using
independent techniques and implementations can now all carry out
reliable simulations of binary black hole coalescence and merger. It
may soon be possible to simulate the black hole binary inspiral
starting from a sufficiently large binary separation so that it can be
compared with and matched to post-Newtonian predictions.  The past
year has indeed seen dramatic progress in numerical relativity
simulations of binary black holes.

{\bf References:}

Alcubierre, M., Br\"ugmann, B., Diener, P., Guzm\'{a}n, F. S., Hawke, I.,
Hawley, S., Herrmann, F., Koppitz, M., Pollney, D., Seidel, E., \&
Thornburg, J., 2005, Phys. Rev. D {\bf 72} 044004.
\hfil\break
Baker, J. G., Centrella, J., Choi, D.-I., Koppitz, M., \& van Meter, J.,
2005, submitted (also 
\htmladdnormallink{gr-qc/0511103}{http://arXiv.org/abs/gr-qc/0511103}).
\hfil\break
Baker, J. G., Br\"ugmann, B., Campanelli, M., Lousto, C. O. \& Takahashi, R.,
2001, Phys.~Rev.~Lett. {\bf 87}, 121103.
\hfil\break
Br\"ugmann, B., Tichy, W., \& Jansen, N., 2004, Phys.~Rev.~Lett. {\bf 92},
211101.
\hfil\break
Campanelli, M., Lousto, C. O., Marronetti, P., \& Zlochower, Y.,
2005, submitted  (also 
\htmladdnormallink{gr-qc/0511048}{http://arXiv.org/abs/gr-qc/0511048}).
\hfil\break
Cook, G. B., \& Pfeiffer, H. P., 2004, Phys. Rev. D {\bf 70}, 104016.
\hfil\break
Diener, P., Herrmann, F., Pollney, D., Schnetter, E., Seidel, E.,
Takahashi, R., Thornburg, J., \& Ventrella, J., 2005, submitted (also 
\htmladdnormallink{gr-qc/0512108}{http://arXiv.org/abs/gr-qc/0512108}).
\hfil\break
Herrmann, F., Shoemaker, D., \& Laguna, P., submitted (also
\htmladdnormallink{gr-qc/0601026}{http://arXiv.org/abs/gr-qc/0601026}).
\hfil\break
Pretorius, F., 2005, Phys. Rev. Lett {\bf 95}, 121101.

\vfill\eject

\section*{\centerline 
{Workshop on Emergence of Spacetime}}
\addtocontents{toc}{\protect\medskip}
\addtocontents{toc}{\bf Conference reports:}
\addcontentsline{toc}{subsubsection}{\it  
Workshop on Emergence of Spacetime, by Olaf Dreyer}
\parskip=3pt
\begin{center}
Olaf Dreyer, Imperial College
\htmladdnormallink{o.dreyer-at-imperial.ac.uk}
{mailto:o.dreyer@imperial.ac.uk}
\end{center}

On the weekend of November 18 the Perimeter Institute hosted a
workshop on the emergence of spacetime. The workshop was organized by
B. G. Sidharth from the Birla Science Centre in Hyderabad, Lee Smolin
from the Perimeter Institute, and Olaf Dreyer, now at Imperial College
London.

The aim of the workshop was to discuss a problem that all quantum
theories have in common: How does a classical spacetime emerge? This
problem of emergence has a technical and a conceptual component. The
technical part is that it is usually very hard to infer details of the
dynamics for a given large quantum system. The conceptual problem is
the added difficulty that arises when a basic concept such as time is
to emerge, as is widely expected to be the case in quantum gravity.
The workshop was conceived to address both these issues.

To shed light on the technical problem we invited solid state
physicists to the workshop. The solid state community has always dealt
with large quantum systems and has developed techniques to describe
their dynamics. This community has in particular stressed the
importance of emergence. Large quantum systems can have properties
that emerge on the level of the whole system but do not make sense on
the level of the constituents. In recent years, members of the solid
state community have started to see this emergent point of view as a
paradigm for all of physics including gravity. Their views then also
provide a new take on the conceptual problem.

In order to allow for in depth discussions the workshop limited the
number of formal talks.  All formal presentations were given on
Friday. The speakers were Grigori Volovik, Renate Loll, Xiao-Gang Wen,
Fotini Markopoulou, Peter Horava, and Seth Lloyd. These presentations
set the stage for the weekend where the atmosphere was much more
informal. The presentation can be found on the website of the
Perimeter Institute 
\htmladdnormallink{http://www.perimeterinstitute.ca}
{http://www.perimeterinstitute.ca}.

The in-depth discussions on the weekend lasted one to two hours and
consisted of a short presentation on the black board followed by a
long set of questions. This format allowed the participants to really
familiarize themselves with the different approaches and see their
advantages as well as their shortcomings. The final overview on Sunday
was unique in that it reviewed all the approaches and listed their
pros and cons. I think this last part was a first in a workshop on
quantum gravity.

The point of view presented by Grigori Volovik posits that the physics
we see around us is described by the ground state of a fermionic many
body system. Such ground states are characterized by the topology in
momentum space. The relevant momentum space topology for us is that of
a Fermi point. These are points on the Fermi surface where the
excitations become gapless. The physics near such a Fermi point is
remarkable in that it looks a lot like current high energy physics.
Lorentz invariance, gauge symmetries and also a dynamic metric are all
emerging.

Renate Loll presented exciting new results in causal dynamical
triangulations. Having worked their way up from two and three
dimensions they have now arrived at four dimensions. The results so
far are promising in that they show the correct dimensionality of four
emerging at large scales. A very curious feature of the approach seems
to be that at Planck scale the dimensionality becomes effectively two.
The significance of this observation is not yet clear.

A presenter that stayed clear of quantum gravity proper was Xiao-Gang
Wen. His presentation focused on the other pillars of our current
understanding of fundamental physics: fermions and gauge interactions.
Wen showed how these objects could emerge from a fundamental theory
made up of simple quantum spins. A ground state of the system called
spin-net condensate has excitations that are fermionic and have
interactions described by a gauge theory.

Fotini Markopoulou described her attempt to deal with the conceptual
problems of quantum gravity. For her the spacetime should emerge from
the interactions of persistent degrees of freedom. To define such
degrees of freedom she introduced noiseless subsystems, a notion
borrowed from quantum information. The persistent degrees of freedom
are noiseless with respect to the evolution of the system.

A connection between solid state physics and string theory was shown
by Peter Horava. The Fermi points introduced by Grigori Volovik also
appear in the physics of D-branes. The formulae describing the
behavior near a Fermi point used by G. Volovik turn out to be a
special case of the Atiyah-Bott-Shapiro construction in K-theory.
Peter Horava proceeded to use these constructions for a new kind of
emergent spacetime in string theory.

A completely new approach to quantum gravity was presented by Seth
Lloyd. His model is based on a quantum computer. He showed how every
quantum computation can be viewed as a superposition of histories and
how every such history can be viewed as a spacetime with matter. The
quantum computation is thus a quantum superposition of spacetimes.

The most interesting outcome of the workshop is the path that a number
of participants have chosen to address the conceptual part of the
emergence problem. They have made progress by assuming a fiducial
time. The interesting question is: Does this step invalidate the
progress that has been made? The final discussion showed that it is
still too early to decide. Crucial steps still remain to be taken. In
the approach presented by Renate Loll recent results have shown that
the dimensionality of the emergent spacetime is correct but it is
still not clear whether gravity is correctly described.

Another thing that became clear during the workshop is that the time
frame for quantum gravity is beginning to change. Quantum gravity
research has been going on for more then sixty years and has not had
much success. With such a time frame the expectations tend to erode
and nobody seems to be rushed. With the results presented by Renate
Loll and Seth Lloyd though this situation seems to change. These
programs might be able to produce quantum theories of gravity in a
time frame of a couple of years rather then decades.

The situation we would be facing then would be a new and welcome one.
Instead of having no theory of quantum gravity we would have several
competing ones. One would then have to decide which one of these is
the correct theory. A task that will require the competing theories to
make observable predictions. What a thrilling prospect.

\vfill\eject

\section*{\centerline 
{Quantum gravity subprogram at the Isaac Newton Institute} }
\addcontentsline{toc}{subsubsection}{\it  
Quantum gravity subprogram at the Isaac Newton Institute, by Jorma Louko} 
\parskip=3pt
\begin{center}
Jorma Louko, University of Nottingham
\htmladdnormallink{jorma.louko-at-nottingham.ac.uk}
{mailto:jorma.louko@nottingham.ac.uk}
\end{center}

The Isaac Newton Institute programme ``Global Problems in Mathematical
Relativity'', which spanned close to 5 months in the autumn of 2005,
contained in October-November a four-week subprogramme on quantum
gravity, organized by Abhay Ashtekar and Piotr Chru\'sciel.  While
quantum issues did feature throughout the mathematical relativity
programme, and especially during the black holes theme weeks in
August-September, the purpose of the quantum gravity subprogramme was
to focus on loop quantum gravity and related topics.

There were two formal and three informal seminars each week. The
formal seminars were pedagogical, targeted at a classical relativity
audience, while the informal seminars were more specialized.
Questions raised at a formal seminar would typically set the agenda
for the next informal seminar or two.  In my perception this
organization worked well in stimulating interaction between
participants from different backgrounds, and the informal seminars
often drew a substantial non-specialist audience.  Outside the
official activities there were numerous informal discussions on
specific topics, and the celebrated layout of the institute building
encouraged all interested to join these discussions. Several postdocs
and research students from the Department of Applied Mathematics and
Theoretical Physics (University of Cambridge) participated in the
activities on a regular basis.

A main theme was the dynamics of loop quantum gravity, including the
mathematical structure of the associated Hilbert spaces and the
concomitant quantization ambiguities. Talks on these topics were
presented by Jerzy Lewandowski, Alejandro Perez, Hanno Sahlmann and
Thomas Thiemann.  Abhay Ashtekar and Martin Bojowald gave talks on
spacetime singularity avoidance in loop quantum gravity, mainly in the
context of quantized cosmological models but with a view to black hole
singularities.  Carlo Rovelli's talk addressed the semiclassical limit
of $n$-point functions in loop quantum gravity.

John Barrett reviewed spin foam models of quantum gravity in three and
four dimensions.  Jorma Louko addressed group averaging techniques in
quantization.

Among the informal seminars, Chris Fewster gave a pedagogical
introduction into algebraic quantum field theory in curved spacetime
and discussed recent work on energy inequalities.  Ian Moss and David
Jennings talked about quantum field effects on accelerated brane
worlds.

The London Mathematical Society organized an afternoon of three talks
aimed at the general mathematical community.  Abhay Ashtekar gave here
an overview of loop quantum gravity, and Karsten Danzmann reviewed the
status of gravitational wave observatories.  Roger Penrose presented a
new perspective on the Weyl curvature hypothesis, suggesting that the
future infinity of a spacetime dominated by a positive cosmological
constant could be conformally reinterpreted as the initial singularity
of a new spacetime.

The subprogramme was intensive and will undoubtedly prove valuable.
It was also fortuitous in overlapping with a Cambridge production of
Carl Djerassi's play ``Calculus'', which dramatizes events around the
Royal Society Committee that passed judgment on the Newton-Leibnitz
priority dispute.  A~number of participants went to see the play;
however, none were to my knowledge among those audience members who
were invited to the stage to become, if only momentarily, members of
the Royal Society.

\vfill\eject

\section*{\centerline {
Global problems in Mathematical Relativity}\\
\centerline{at the Isaac Newton
  Institute}}
\addcontentsline{toc}{subsubsection}{\it Global
 problems in Math Relativity at the Newton Institute, by Jim
 Isenberg} \parskip=3pt
\begin{center}
Jim Isenberg, University of Oregon
\htmladdnormallink{jim-at-newton.uoregon.edu}
{mailto:jim@newton.uoregon.edu}
\end{center}

Appropriately coinciding with last year's centenary of Einstein's
great papers, the Isaac Newton Institute in Cambridge, England,
sponsored and hosted during 2005 a nearly 5 month long programme on
"Global Problems in Mathematical Relativity". The programme (organized
by Piotr Chrusciel, Helmut Friedrich, and Paul Tod) was remarkably
rich, extensive, and varied. Included were weeks of concentration on
each of the following topics:

--the analysis of hyperbolic PDEs, including Einstein's equations

--numerical relativity

--black holes

--Einstein's theory as a dynamical system

--applications of Riemannian geometry in general relativity

--applications of Lorentzian geometry in general relativity

--global analysis and global techniques

--quantum aspects of gravity

--asymptotic structures in general relativistic  spacetimes

--the application of inverse scattering techniques to the studies of  solutions of Einstein's equations

--static and stationary solutions

--the Einstein constraint equations and their solutions

Each of these concentration periods attracted researchers from all
over the world. In addition to the 7 mathematical relativists plus 2
graduate students who were there for the entire programme from early
August until the end of December, there were roughly 10 to 15 shorter
term visitors during any given time. With a light schedule of 3 or 4
talks per week, the emphasis was on concentrated collaborations among
the participants. It was not unusual to see intense black board
sessions occurring at all hours from 7 in the morning until well past
midnight. By last count, at least 26 papers submitted for publication
have resulted from collaborations carried out during the programme.

In addition to the weekly schedule of talks included in the programme,
there were 4 special conferences. One of them was held as a satellite
meeting at Southampton University. It focussed on numerical
relativity, and reported on some of the breakthroughs for binary black
hole simulations that have happened just this past year.  The other
special conferences were held at the Newton Institute. The first of
these, the week long Euroconference on "Global General Relativity",
included talks on a very wide range of topics, from recent
developments on quasilocal mass to the latest observational data
pertaining to astrophysical black holes, and from numerical
simulations of classical solutions to recently developed ideas on
quantum field theory in curved spacetimes. This conference was very
popular, attracting over 100 participants. Equally popular was the one
day "Spitalfields Day" (co-sponsored by the London Mathematical
Society), which consisted of three lectures on the general theme of
"Einstein and Beyond". These less technical lectures, delivered by
Abhay Ashtekar, Roger Penrose, and Karsten Danzmann, attracted
standing-room-only audiences for discussions of gravitational
radiation detection, quantum gravity, and the nature of the universe
at late times. The five month long Newton Institute programme was
capped by a week long Euroconference in December which focussed on
studies of the Einstein constraint equations and on a number of
related mathematical and physical themes. This conference particularly
highlighted the very important symbiotic relationship between
geometrical analysis and mathematical relativity.

In addition to publicizing some of the particular recent triumphs
which have occurred in mathematical relativity (including gains in
understanding the nature of gravitational fields near cosmological and
black hole singularities, as well as the rapid development of powerful
new techniques for studying the stability of black hole spacetimes),
and in addition to providing a perfect environment for the development
of collaborations among workers whose home bases are widely scattered
around the globe, the Newton Institute programme served as an
important notice to the community of mathematicians and physicists
that mathematical relativity is a very healthy discipline, which has
had a great impact on both physics and mathematics, and will continue
to do so. As the programme broke up just before the end of the
Einstein centenary year of 2005, the participants all hoped to soon
have another opportunity to collaborate and to share ideas in such an
ideal setting.

\vfill\eject

\section*{\centerline 
{Loops '05}}
\addcontentsline{toc}{subsubsection}{\it  
Loops '05, by Thomas Thiemann}
\parskip=3pt
\begin{center}
Thomas Thiemann, Albert Einstein Institute
\htmladdnormallink{thiemann-at-aei.mpg.de}
{mailto:thiemann@aei.mpg.de}
\end{center}

In the Einstein year, the almost annnual conference on background
independent approaches to quantum gravity took place at the Albert
Einstein Institute in Potsdam, close to Berlin, Germany. The official
title of the conference was `Loops 05', however, not only Loop Quantum
Gravity (LQG) researchers were present but also practitioners of the
other non -- perturbative approaches. Plenary talks were distributed
among the following topics: 1. Asymptotically Safe Quantum Gravity, 2.
Causal Sets, 3. Dynamical Triangulations, 4. Generally Covariant
Algebraic Quantum Field Theory and 5. Loop Quantum Gravity.  Almost
all the leaders in those fields were present at the conference, in
particular, 1. Reuter, 2. Sorkin and Dowker, 3. Loll, 4.  Verch, 5.
Ashtekar, Baez, Barrett, Corichi, Freidel, Gambini, Lewandowski,
Perez, Pullin, Rovelli and Smolin. There were also talks on background
independent aspects of string theory (Dijkgraaf and Theisen),
Supergravity (Julia), Emergent Quantum Gravity (Morales -- Tecotl) and
Quantum Cosmology (Maartens).

There were 20 plenary talks and 63 afternoon talks which, for the
first time, had to be distributed over two parallel sessions. We had
more than 150 official registrations but the lecture theater was
sometimes filled close to capacity (210 seats). This was certainly the
biggest quantum gravity conference focusing on background independent
approaches so far. It is pleasing to observe that the number of
participants at this kind of meetings is rapidly increasing. From my
own memory I recall the following conferences and rough participant
numbers respectively: Banach Center, Warsaw, Poland, 1995 (50); Punta
Del Este, Uruguay, 1996, (40); ESI, Vienna, Austria, 1997, (60);
Banach Center, Warsaw, Poland, 1997 (60); ITP, Santa Barbara, USA,
1999, (70), Banach Center, Warsaw, Poland, 2001 (60); IGPG, State
College, USA, 2003 (90); CPT, Luminy, France, 2004 (110).

The conference was subsidized by the Max Planck Gesellschaft (MPG) and
The Perimeter Institute for Theoretical Physics (PI). While PI
sponsored the conference poster, the money from the MPG and about 60\%
of the conference fee (EUR150,-), which was cashed only from non --
students, was solely used in order to enable students to participate.
I would like to take the opportunity to thank the plenary speakers
once again for not asking for reimbursement which would have down-sized
the student participation by an order of 40 people.

Due to the help of the marketing company `Milde Marketing', the
conference also had quite some impact on the German press. Major
articles appeared for instance in the Frankfurter Allgemeine Zeitung
and the television company RBB interviewed some of the participants
and intends to broadcast parts of the conference. Also, Lee Smolin
spoke in the `Urania', a world famous institution in Berlin, which
focuses on mediating science to the public through popular talks.

The scientific contributions to the conference can be downloaded, in
many cases both audio and video, from the conference website
\htmladdnormallink{http://loops05.aei.mpg.de}{http://loops05.aei.mpg.de}.
It is difficult to single out particular highlights but maybe one of
the lessons to take home from the conference is that all afore
mentioned approaches start deriving results relevant for quantum
cosmology which is very important in view of the fact that precision
cosmological measurements such as WMAP and later PLANCK might be able
to detect quantum gravity fingerprints in the CMB.

Personally, I would be very pleased if somebody took the energy to 
organize Loops06: Let us keep up the momentum and make this meeting an
annual forum of our small but growing community.  

\vfill\eject

\section*{\centerline 
{Numrel 2005}}
\addcontentsline{toc}{subsubsection}{\it  
Numrel 2005, by Scott Hawley and Richard Matzner}
\parskip=3pt
\begin{center}
Scott Hawley and Richard Matzner, University of Texas at Austin
\htmladdnormallink{matzner2-at-physics.utexas.edu}
{mailto:matzner2@physics.utexas.edu}
\end{center}

This workshop was organized by Dr. Joan Centrella, the Leader of the
Gravitational Astrophysics Laboratory of the Exploration of the Universe
Division at NASA/Goddard. The presentations at the workshop are available on
line at: 
\htmladdnormallink
{http://astrogravs.gsfc.nasa.gov/conf/numrel2005/presentations/}
{http://astrogravs.gsfc.nasa.gov/conf/numrel2005/presentations/}

The workshop offered attendees an
excellent overview of cutting-edge research throughout the field.
Representatives from most of the major numerical relativity groups
(AEI, Austin, Baton Rouge, Brownsville, Caltech, Goddard, Penn
State...) were present and presented new research results.
We heard about remarkable new progress in being able to simulate binary
black hole interactions and to extract waveforms. Some of this work (by
Pretorius, and a collaboration including Pollney and Diener) had been
previously presented and/or posted at gr-qc, but two of the presentations
(by Zlochower, representing a collaboration led by UTB, and by Choi
representing a collaboration led by NASA/Goddard) had been previously
unpublished in any form.

The opening was given by Dr. Nick White, the Director of the Exploration of
the Universe Division at NASA/Goddard. We immediately went into a
presentation by Joan Centrella: ``Gravitational Wave Astrophysics, Compact
Binaries, and Numerical Relativity". Centrella began with a discussion of
the fundamental aspects of gravitational radiation. She followed with a
discussion of the sensitivity of the LIGO detectors, which are at
essentially full sensitivity as the Science Run S5 begins. Real
gravitational wave science has begun.  Even higher sensitivity (in a much
lower frequency band) is expected for the proposed space-borne detector LISA.

What are the expected sources for gravitational wave detectors? LISA will be
sensitive to supermassive black hole mergers. ``Every" galaxy has a central
supermassive black hole; ``every" galaxy has undergone a merger. ``X-type"
radio sources (in disturbed galaxies) show sudden changes in jet direction,
which is assumed to lie along the spin direction; a merger could lead to a
flip of the spin, producing such a change, so these are considered
supporting evidence for the existence of mergers. For LISA and (perhaps) for
LIGO, intermediate mass black holes (hundreds of solar masses) are possible
sources of gravitational radiation. Here there is fairly strong evidence
from X-ray sources in active star forming galaxies. Stellar mass black holes
can result from the explosion of high mass stars, and there is a chance that
binary black holes can form by this method. This is of course the underlying
assumption in computational gravitational waveform prediction from binary
black hole mergers. Finally, there is recent evidence, from observations of
the short gamma ray bursters GRB 050509b, GRB 050724, suggesting they are
mergers (neutron star/ neutron star or neutron star/black hole) that leave a
black hole which shows evidence of accretion-driven X-rays after the merger.
These could be sources for Advanced LIGO.

Deirdre Shoemaker discussed ``Complexity in Gravitational Waveforms from BH
mergers". Her point was that in fact we are seeing little complexity. How do
we understand the relative similarity of all
gravitational waveforms obtained via numerical simulations? Shoemaker put
forth a few ways in which ``complexity" could be added to (and ultimately of
course, extracted from) the waveforms. Shoemaker began by noting that there
is beginning to be a strong convergence of results in binary black hole
simulations, at least for simple configurations. She presented head-on
simulations from Zlochower et al, Fiske et al., and from Sperhake et al. The
waveforms are very similar, and are rather smooth with a rapid onset to
apparent ringdown behavior. For the existing orbit-and-plunge simulations,
the agreement is not especially tight. However they too show fairly smooth
behavior and fairly rapid onset of ringdown.

Shoemaker raises the question: why is there so little complication in the
waveforms? She points out that this is in contrast to the case with matter
present, as in neutron-star binaries. She gives several examples (Duffert et
al., Faber et al., Duez et al. ) which have more structure near the end of
the merger than occurs in the binary black hole case. A core collapse
waveform shows quite complicated behavior (Zwerger and Mueller). Shoemaker
presented one aspect of complexity in nonequal mass head-on collisions.
``Kicks" occur in these cases, and the waveforms have somewhat more
structure, but still show early onset of ringdown.

In order to examine the mechanism of early ringdown onset Shoemaker
considered scalar field ringdowns, carried out in a sequence of fixed
spacetimes which constitute a sequence of quasi-circular initial data
(Pfeiffer et al 2004). The ringdown is Fourier transformed, and the
frequency identified in the scalar ringdown. For a single black hole the
frequency of the ringdown is related to the mass (in Schwarzschild) by
$ M_{BH}= 0.29293/\omega$. In these data, the {\it ADM} mass measures the
total system mass, roughly twice the horizon mass of one hole.

If the holes are well separated (corresponding to a time well prior to the
merger in an evolution), $M_{BH}$ determined from the ringdown will be the
mass of one of the two equal masses. As one considers holes close enough
together (in principle corresponding to later time in the merger), the
system acts as if it has a single effective ringdown potential with a mass
equal to the ADM mass.  For her survey Shoemaker finds the transition occurs
roughly at the ``ISCO", with a horizon separation of $8m. $ This again
support s the idea the merger forms a common potential and quickly moves
into the final black hole ringdown, fairly early in the evolution (when the
ISCO is reached).

Yosef Zlochower, from The University of Texas at Brownsville,
discussed ``Accurate Binary Black Hole Evolutions Without Excision".
This previously unpublished work caused quite a stir. Zlochower gave
an introduction to the BSSN method with punctures. In the puncture
method, one sets data for black hole evolution by using modified
Brill-Lindquist data. These data have $r^{-1}$ singularities in the
conformal factor, at the coordinate centers of the black holes.
Previous to this workshop, known approaches either used excision, to
cut the singular region out of the computational domain (obviously not
specific to puncture data), or held the punctures fixed at their
initial coordinate positions, treated the singular parts analytically,
and computationally evolved only the subdominant nonsingular part of
the conformal factor. (Work by Bruegmann and collaborators, reported
at the workshop by Tichy, uses corotating coordinates to evolve a
black hole binary for one orbit in a system where the punctures are at
fixed coordinate position, though Bruegmann {\it et al.} also
considered excised evolution.) However Zlochower then showed inspiral
and merger results in which the full punctures were evolved and
traveled across the computational grid! This was so unexpected that it
led to a flurry of questions, and many points were only partially
explicated before the next presentation had to start. However the work
has now been posted on arXiv.org:
\htmladdnormallink{gr-qc/0511048}{http://arxiv.org/abs/gr-qc/0511048}.
That publication explains that the code is designed to evolve the
inverse of the conformal factor. That, and some care to the behavior
of the lapse, allows evolution of the full system, including the
``singular" conformal factor. Second order convergence is shown. The
data correspond to late inspiral. The holes perform about half an
orbit before a common apparent horizon forms. A waveform is extracted,
and the energy radiated is of order 2.8\%, argued to be accurate to
about 10\% of this value. About 14\% of the total angular momentum was
radiated. The domain represented in the evolution extends to
approximately 60M (where M is the total mass), via the use a a
multiple step ``fisheye" transformation.

The next presentation, by Dae-Il Choi from Goddard Space Flight
Center: ``Gravitational Waveforms from Coalescing Binary Black Holes",
presented very similar results by very similar methods. This work has
also been posted on arXiv.org:
\htmladdnormallink{gr-qc/0511103}{http://arxiv.org/abs/gr-qc/0511103}.
(Apparently the two groups were unaware of the others' work until
these presentations.) Choi's presentation described a numerical
regularization of the conformal factor singularity, which does not
involve introducing the inverse of the conformal factor. Instead,
straightforward finite differencing, together with a careful choice of
the lapse allows the evolution of the puncture system. The initial
data put the holes, and hence the punctures, on the z = 0 plane. The
Goddard code uses a cell centered formulation, so no point at which
quantities are evaluated actually contains this plane. For nonspinning
holes, the orbits stay in the $z=0$ plane and thus every computed
quantity is finite. The Goddard code uses fixed mesh refinement with
eight levels of factor of two steps. The outer boundary is at $128 M$,
and the inner box resolution in three different resolution simulations
is $M/16$, $M/24$, and $M/32$. Second order convergence is shown. The
simulation yields a waveform and the total radiated energy is
$0.0330M$, where $M$ is the total mass of the system. The angular
momentum radiated is $J=0.138 M^2$. Further, the radiated energy was
computed in two ways. The first computation was by integrated
gravitational radiation flux across a sphere at radius equal to $25$
times the horizon radius of the final black hole. Then this was
compared to the difference of the initial and final ADM mass. The loss
of mass-energy closely checks between these two methods. In fact the
two measures of the wave energy loss agree to better than 5\% (of the
$\approx 3\%$ of the total energy that is radiated), showing at least
$3-$digit accuracy.

Alessandra Buonanno of the University of Maryland discussed ``Predictions
for the last stages of inspiral and plunge using analytical techniques".
She described a method (the {\it Effective One Body} (EOB) method), to resume
post-Newtonian expansions, in a way that appears to converge better than the
straightforward PN expansions. Within analytical calculations the EOB is the
only method which
can approach a description of the dynamics and the gravity-wave signal
beyond
the adiabatic approximation. It can also provide initial data ($g_{ij}$,
$K_{ij}$) for black holes close to the plunge to be used by numerical
relativity. The real difficulty in carrying out these processes lies in
understanding how accurate the EOB method really is, since it is an
expansion in a PN-like parameter. At some point must we use more accurate
(presumably computational) description.
Current results indicate good agreement between numerical and
analytical estimates of the binding energy without spin effects.
Already in its current state the method can be used as a diagnostic for (or
to fit) numerical relativity results, and can also suggest parameters to
vary in templates to provide more complete coverage of the space of possible
waveforms.

Wolfgang Tichy (Florida Atlantic University) gave a discussion of
``Simulations of orbiting black holes" This work is a review of work done by
Tichy, Jensen and Bruegmann ({\it Phys. Rev. Lett.} {\bf 92} 211101 (2004);
\htmladdnormallink{gr-qc/0312112}{http://arxiv.org/abs/gr-qc/0312112}). 
This approach uses puncture initial data for two orbiting black holes, with
a (fairly standard) modified BSSN (Baumgarte Shapiro Shibata Nakamura)
evolution system which replace all undifferentiated $\tilde \Gamma$ by
derivatives of the metric, and ensures the tracelessness of the traceless
conformal extrinsic curvature by explicit subtraction of the trace of
$\tilde A_{ij}$ from $\tilde A_{ij}$ after each time step. (The time
integration is Iterated Crank Nicholson).
The outer boundary is a ``lego sphere", with Sommerfeld outer boundary
conditions for all evolved quantities.

The evolution is run with the holes excised (comparable results were
found with puncture evolution that analytically handled the singular
part of the conformal factor). The excision was carried out on lego-spheres
of the black holes inside the horizon by the process of
copying the time derivative at next interior point onto excision
boundary (simple excision). Singularity avoiding gauge (lapse) was
used to prevent the slice from running into physical singularities.
The code is based on the BAM infrastructure, using fixed mesh
refinement (FMR) for efficiency. Seven levels of nested boxes were
constructed around each black hole. Outer boundaries were compared at
24M, 48M, and 96 M. The code evolves equal mass nonspinning holes, and
uses quadrant symmetry.  Because of these settings, the code can be
run on a laptop. Perhaps the most important innovation was the use of
co-moving coordinates enforced via the shift vector, which compensate
for black hole orbital motion. The dominant terms in the shift
correspond to rigid rotation. However it is the case that the data do
not describe exactly circular motion, so the black holes can drift
from their coordinate location in a specific evolution. An interesting
algorithm is used to modify the shift to center the value of the lapse
function (a proxy for the apparent horizon location) at a fixed
specific coordinate location. The evolutions can evolve up to about
125M, more than the orbital time-scale inferred from the initial data.
However, no waveforms have been published from these runs.

Peter Diener (LSU) discussed ``Recent developments in binary black hole
evolutions". The results were obtained with a large group of collaborators
(M. Alcubierre, B. Bruegmann, F. Guzman, I. Hawke, S. Hawley, F. Herrmann,
M. Koppitz, D. Pollney, E. Seidel, R. Takahashi, J. Thornburg and J.
Ventrella) associated with the Max Planck Institute Albert Einstein
Institute in Potsdam, Germany, and with LSU.

This is another example of the surprising gains that have been made in
computation of binary black hole interactions. Diener presented work using
fixed mash refinement, with corotating coordinates, and an active adjustment
of the shift vector to keep the holes centered in the corotating
coordinates. Developing this code was prompted by the earlier work by
Bruegmann,  Tichy and Jansen, reported above by W. Tichy. Very
interestingly, in this code the gauge used had five adjustable parameters,
which control the behavior of the lapse and the shift. The lapse and shift
are determined by driver conditions that evolve them toward ``1+log" lapse
and co-moving shift (which follows the centers of the black holes). These
parameters define the factor of the lapse in the time derivative controlling
the shift, for instance.

Two sets of parameters were used: {\it Gauge Choice 1}, and {\it Gauge
Choice 2}. It was found that the lifetime of the simulation depended on the
gauge used (Gauge choice 1 ran longer, to about 140M; Gauge choice 2 ran  to
about 80M). Also the computed proper distance (at a given time) between the
apparent horizons depended on the gauge choice. When the two gauges were run
at various resolutions, however, one could do convergence on the proper
separation, and Richardson extrapolation. The result is that the
extrapolated separations from the two sets of simulations overlap very
closely in the range where both are available ($t$ < $80M$). This is a very
interesting result. It means that the code is definitely doing something
right. It strongly suggests that the different implementations of the driver
conditions are leading to very closely the same time slicing. It also
indicates that quite fine discretization is needed to achieve the convergent
regime, and to achieve reasonable accuracy.

Denis Pollney of the Albert Einstein institute in Potsdam, spoke on
``Evolutions of Binary  Black Hole Spacetimes in the Last Orbit" He broke
his talk into two sections: 1. Evolutions of Helical Killing
Vector Data; 2. Evolving Single Black Holes Using a Multi-patch Code.

In the binary orbit (helical Killing vector) work, comparisons were
made of evolutions from data set in one of two ways: Punctures with
parameters along an effective potential sequence developed by Cook
(1994), and thin sandwich data using a {\it Helical Killing Vector}
condition, constructed by Grandclement, Gourgoulhon and Bonnazolla
(2002) Meudon data. This code is as described in Diener's talk above;
in particular it is a rectangular coordinate code. A series of orbits
and head-on collisions can be produced in this code, and in particular,
results similar to those of the earlier work by Bruegmann, Tichy and
Jansen were accomplished.

The second aspect of Pollney's presentation concerns patching to
spherical domains, and in particular conforming surfaces for inner
(excision) and outer boundary surfaces. The current work has
concentrated on single black holes, and uses Thornburg's inflated cube
coordinate system, which is multi-patch (six patches) and uses
interpolation between adjacent grids.  Angular coordinates are chosen
so that adjacent patches share coordinates perpendicular to their
mutual boundary so the method needs only 1D interpolations. The method
was tested in a hydrodynamics code (Whisky) to show that shock
propagation is correctly handled across the interfaces in a single
hole background. Evolution of a single distorted black hole by this
method showed inter-patch effects well below finite difference
accuracy.

Frans Pretorius of the University of Alberta described his binary
black hole simulations. [{\it Phys. Rev. Lett.} {\bf 95} 121101
(2005);
\htmladdnormallink{gr-qc/0507014}{http://arxiv.org/abs/gr-qc/0507014}]. 
This work integrates several approaches which
are not in wide use within the numerical relativity community --- use
of a second-order formulation of the Einstein equations,
compactification of spatial infinity, adaptive mesh refinement, and a
harmonic gauge. The code is based on generalized harmonic coordinates,
so every component of the Einstein equations obeys a wave equation
(with nonlinear source).  Source functions are added to the definition
of the harmonic gauge in order to be able to choose slicing and shift
conditions. These equations are discretized directly into second-order
in time finite difference equations. Adaptive mesh refinement follows
{http://www.perimeterinstitute.ca}the holes and excision removes the singularities from the domain.
Numerical dissipation is used to control instabilities that otherwise
arise near the black holes.  A final interesting point is that the
spatial domain is compactified, which provides an ``inexpensive''
implementation of the outer boundary. In addition, Pretorius used {\it
  constraint damping} which adds a linear combination of the gauge
constraints to the metric evolution, and which produces extended
stable evolution. Initial data are set up using boosted field
collapse. The initial data slice is conformally flat maximal.  The
harmonic condition gives the initial time derivatives of the lapse and
shift. The Hamilton and momentum constraints are solved for the
initial conformal factor and shift.

Data as set are equal mass components, in an approximately circular orbit
(eccentricity of order 0.2 or less), with a proper distance between holes of
approximately 16M (coordinate separation of 13M), where the initial ADM mass
is 2.4M. Each black hole has a velocity of abut 0.16, and zero spin angular
momentum.
The evolution traces out about 2/3 of an orbit to one full orbit before the
horizons merge. The final black hole has an angular momentum Kerr parameter
$a=0.7M_f$. Presumably because of dissipation and lack of resolution in the
spatial compactification, the extracted waveform has a faster than $r^{-1}$
falloff, though the shape is very well preserved at different wave
extraction radii. Regardless of the falloff, one estimates about 5\% of the
total initial energy is radiated.

Mark Scheel (CalTech) discussed ``Solving Einstein's Equations using
Spectral Methods".  He began with a description of the method, which
provides exponential convergence as the number of basis functions is
increased (though the {\it coefficient} in the exponential must be studied
in each case). In the pseudospectral approach, analysis of the basis
functions leads one to define specific collocation points $x_n$. Expansions
to truncated series in terms of basis amplitudes and basis functions
(truncated to a maximum number, $N$) can be inverted exactly to obtain the
basis amplitudes by carrying out a weighted sum with $N$ terms, over the
function at the specified collocation points, multiplied by the basis
functions at those points. This transformation between space and spectral
representation is an algebraic process. In the nonlinear case derivatives
are computed in spectral space, nonlinear terms are evaluated in physical
space. Boundary conditions are imposed analytically on characteristic
fields. 

The method uses  characteristic decomposition, and complicated domain
decomposition; every domain maps either to a spherical region or to a
sphere, where the basis functions are defined. An example for a single black
hole had 54 sub-domains. Because of the fact that the sum over basis
functions defined a function everywhere, no explicit interpolation is needed
to transfer values between patches.

The KST code [Kidder, Scheel, Teukolsky, {\it Phys. Rev.} {\bf D} 64 064017
(2001)] is a parameterized hyperbolic code. This was used with
``quasi-equilibrium" conformal thin sandwich data to compute binary black
hole interaction in a co-moving frame. By using characteristic
decompositions, no boundary conditions are needed at the BH horizons
(excision) and Sommerfeld-like outer boundary conditions were imposed at
the outer $r= 320M_{BH}$. The was a free evolution, the constraints were
solved only initially. For moderately short evolutions the constraints
converged. However by $t \approx 20M$ the convergence was lost. A fix in the
shift vector to keep the apparent horizons centered in coordinate location
(and spherical) improved the behavior for about another $10M$, but
convergence was then lost. Suggested fixes were to impose elliptic gauge
conditions, or some sort of driver condition.

The final topic of Scheel's talk concerned an effort to construct a
pseudospectral version of Pretorius code, necessarily adapted to a first
order form, this code works extremely well for single Schwarzschild black
holes. However, very strong instabilities are found when trying to do
co rotation problems. with moderately large domains. Even flat space is
unstable (when $ R\Omega > 0.7$, with $R$ the size of the domain an $\Omega
$ the angular velocity). The KST system does not have this problem.

Harald Pfeiffer (CalTech) discussed ``Quasi-equilibrium binary black hole
initial data". The basic idea is that there is approximate time independence
in a corotating frame; this implies an approximate helical Killing vector.
Time-independence in corotating frame
implies vanishing time derivatives.
The idea is that the initial data for black holes in not too close orbit
approximately satisfy these condition; construct data that exactly has these
properties. The solution proceeds by a conformal solve of the resulting
elliptic equations to obtain the lapse, the conformal factor and the shift
vector The co rotation requires a boundary condition on the shift vector of
$\beta^i = (\omega \times r)^i$. The boundary condition on the lapse at
infinity is  $N=1$, and on the conformal factor $\psi = 1$.
Inner boundary conditions on $\psi$ and $\beta$ are written at the apparent
horizons, which are assumed in the data to be stationary and isolated (no
shear of their generators). The so called
extended conformal thin sandwich formalism
also sets the time derivative of the extrinsic curvature on the initial
slice to zero. This formalism leads to some curious double valuedness in the
ADM energy as a function of wave amplitude in the conformal background. This
apparently can be understood physically, and can be evaded by considering
only the standard conformal thin sandwich approach.

One possible difficulty is that evolution codes typically evolve inside the
apparent horizon, but these data are produced with the apparent horizon as
its inner boundary.

Greg Cook (Wake Forest)gave an update  ``Black-Hole Binary Initial Data:
Getting the Spin Right". This was an update on the construction of
quasi-circular binary black hole data. All of his results made use of the
conformal thin-sandwich method. There are two approaches that yield a
sequence of quasi-circular orbits. In one approach, the binding  energy is
plotted vs the orbital angular momentum, with fixed total angular momentum.
The minimum of each of these curves defines an effectively circular orbit.
The second approach makes use of the fact that in true quasi-circular motion
all the fields should be constant in the corotating frame. one compares the
value of the Komar mass (defined only for stationary spacetimes) with the
ADM mass. Consider the corotating Black Hole case. By computing a number of
test cases, Cook found a modification of the lowest-order corotating
condition for the tidal field seen by a black hole at its horizon, in the
presence of a second:
$\beta^i = \alpha\psi^{-2} \tilde s^i + \Omega_{BH} \xi^i$
at the horizon, where $\tilde s^i $ is the spacelike normal to the
horizon, and $\xi^i$ is a spatial unit vector. The first term $(
\alpha\psi^{-2} \tilde s^i)$ is well established; it is an outward directed
shift component that counteracts the inward fall of the coordinates. The
second term had previously been taken as the angular rate $\Omega_0$
measured at infinity. However Alvi (2000) pointed out that the rate at the
horizon of the tidal rotation is given by

$\Omega = \Omega_0 - \eta M/b + ...$ ,
where $b$ is a measure of the separation, $\eta=m_1m_2/M^2$ , and $M$ is the
total mass. Cook expresses this completely in terms of the rotation rate and
finds $\Omega = \Omega_0 - \eta (M \Omega_0)^{2/3} + ...$ . With this
correction for co rotation, Cook finds complete agreement between the helical
Killing Vector and the effective potential methods.

Scott Hawley (University of Texas, Austin) gave a summary of some recent
work (with Richard Matzner and Michael Vitalo) validating an efficient
multigrid-with-excision code that produces binary black hole data. The code
is a node centered code, and uses a particular way of defining the excision.
The excised points are those on any grid which lie inside the excision
radius. Consequently, except for very special choices of the parameters, the
excised regions are {\it larger} on the finer grids. The code is
parallelized, and exists in a fixed-refinement version. However Hawley spoke
about the unigrid code. To test the code, Hawley compared the computed
binding energy to predictions of a lowest-order spin-spin coupling due to
Wald. For relatively close placement of the momentarily stationary holes in
the initial data set (coordinate separation of $10m$, where each hole has
mass parameter mass $m$), one obtains a binding energy variation with spin
that has the dependence on angle suggested by Wald, and an amplitude about
10\% higher. This latter difference is attributed to the closeness of the
holes in these runs. (The binding energy is computed by assuming the horizon
area determines the intrinsic mass, and subtracting that from the ADM mass.)
More exploratory work will be carried out in the future, with the FMR
version of the code.

Stu Shapiro (Illinois), ``Binaries Containing Neutron Stars: The Merger
Aftermath",  described a number of results concerning neutron stars, and
neutron star/neutron star and neutron star/black hole mergers. He began by
showing computations indicating that certain rotating hypermassive stars are
dynamically stable, but other physics (turbulent viscosity, magnetic
braking, neutrinos/ gravitational waves) can lead to delayed collapse to
black holes and delayed gravitational wave bursts.
Shapiro described simulations of neutron-star binary systems mass ratio
$0.9$ to $1.0$. The work found a critical mass approximately $2.5$ to $2.7
M_{solar}$ for the merged star. Exceeding this critical mass leads to prompt
collapse; less than the critical mass leads to an hypermassive remnant and
delayed ($100msec$) collapse. The associated gravitational wave frequency is
in the $3$ to $4 kHz$ range, a possible target for AdLIGO.
A theoretical question associated with rotating collapse concerns whether
data with $J/M^2 > 1$ can collapse beyond the neutron-star stage.
Computational experiments, some described by Shapiro, do not do so. (Note
that a Kerr solution with  $J/M^2 > 1$ has a naked singularity.) The
behavior of the matter in the simulations is a rotation induced bounce.
Simple Newtonian arguments explain the results by angular momentum
conservation preventing collapse to within a horizon.
Shapiro also described new, more realistic simulations involving General
Relativistic MHD, and including realistic shear viscosity. These effects may
contribute to delayed collapse with a gravitational wave burst, enhanced
collapse bounce shocks, and possible magnetic jets. These topics are some of
the most 
astrophysically relevant things which computational physicists can
approach, and suggest exciting AdLIGO connections: coincident (triggered)
detection between GRBs and their associated gravitational radiation, with a
reasonable event rate.

John Baker gave a summary and``future directions" talk Friday
afternoon.

The work in this conference that produced waveforms (Zlochower, Choi and
Pretorius)  produced waveforms that are remarkably similar in general
features. In particular, most of the waveform``looks like" a ringdown
waveform, and this is where most of the energy is radiated. It does appear
that we can define a``generic" waveform for black hole mergers, appropriate
to template generation. One of the outcomes of the meeting was a brief
meeting of an ad-hoc committee chaired by John Baker to define data standards
for, and to collect, waveform data from simulations, ensure consistency with
the standards, and to post them at a public website.

\vfill\eject

\section*{\centerline {
Apples With Apples Workshop in Argentina}}
\addcontentsline{toc}{subsubsection}{\it  
Apples With Apples Workshop in Argentina, by Sascha Husa}
\parskip=3pt
\begin{center}
Sascha Husa, Friedrich Schiller University Jena
\htmladdnormallink{sascha.husa-at-uni-jena.de}
{mailto:sascha.husa@uni-jena.de}
\end{center}

The third ``Apples with Apples'' workshop, which took place from March 14--25 
2005 in Argentina, continued a series of roughly one 
two-week meeting per year to bring
together numerical relativists in hands-on comparisons of formulations
of the Einstein equations for numerical relativity.
The meeting was  organized by Oscar Reula
in Villa General Belgrano, located in the beautiful
Calamuchita Valley near Cordoba.
The conference hotel that hosted all participants   
provided a very communicative
setting for our purposes. Special thanks go to Oscar and the local students
Florencia
Parisi and Santiago Gomez for their help and support of the participants.
The meeting followed the established patterns of previous apples with apples
meetings, with talks and discussions in the first week, and working sessions
and more discussions in the second week.
Talks have been presented by Jeff Winicour, Osvaldo Moreschi, Tilman Vogel,
Sascha Husa, Santiago Gomez, Bernd Reimann, Carles Bona, Bela Szilagyi,
Yosef Zlochower and Pedro Maronetti, and all of these talks have
been accompanied by rather lively discussions.

Jeff Winicour opened the meeting with a general introduction to the ideas and history of
the  project, and set the scene for the discussions to come.
Carles Bona, Bela Szilagyi and Yosef Zlochower presented
test results with their codes (the Z4 system, different versions of harmonic codes,
and the LazEv BSSN code), and what they had learned from their tests and the discussions within the project. Sascha Husa presented results obtained with 
Calabrese and Hinder in [1] on second order in space hyperbolic evolution
equations, and presented suggestions for revising the robust stability test.
Bernd Reimann (see [2]) and Tilman Vogel (see [3])
discussed their promising approaches to deal with continuum instabilities.
Pedro Marronetti presented his thoughts on setting up tests for binary neutron
star evolutions, followed by a discussion on what could/should be done
regarding tests for systems with matter.  Santiago Gomez presented work of the
Cordoba group on a new evolution system using components of the Weyl tensor as
evolution variables, Osvaldo Moreschi talked about a new approach to the binary
black hole problem, where interior and asymptotic region are matched with
analytic methods. See [4] for abstracts and some slides.

The declared goal of the apples with apples project is to develop a hierarchy of testbeds which should eventually include binary black hole problems, and a natural hope is to progress rather quickly from the simple toy problems with periodic boundaries we had designed at the first meeting to actual black hole spacetimes -- in particular since running 
3D black hole simulations with advanced technology such as grid refinement or excision has become routine for several groups. However, another declared goal of this project is to significantly improve our level of actual understanding -- 
believing that understanding is key to eventually develop robust simulation methods.
More than for the previous meetings, the spirit of the Cordoba meeting has been one of
digestion rather than accelerating the broadening of our scope -- but this, I believe, has been achieved rather successfully!
One of the main topics of the workshop was to incorporate recent theoretical progress into our practical program of designing test suites and in particular also into the interpretation of test results. Most notable here are 
the advances regarding the mathematical understanding of second order in space systems, of continuum instabilities
(e.g. as signified in the talks of Reimann and Vogel), and in much work on particular evolution systems which has directly emanated from the apples project, such as the 
rather detailed studies of the Pittsburgh group.

Let me select a few points where our understanding has improved substantially:
The instability exhibited for the ADM [5] system in the first ``apples paper'' [6] has 
finally been nailed down as an ordinary
von Neumann instability. In order to properly understand this,
progress with the theory of well-posedness and 
numerical stability for second order in space systems was required, see e.g. [1,7].
In fact, one of the misleading original ideas was to look for
exponential growth in our ``robust stability test'', whereas
weakly hyperbolic equations should be expected to only produce
resolution dependent polynomial, e.g. linear, growth in the
linear constant coefficient case (i.e. the robust stability test setup).
This non-convergent behavior has in particular been verified for the ADM system.

Some confusion had been caused by the fact that numerical stability tests in the linear constant coefficient regime can show
a rather complicated phenomenology due to the frequency dependent damping effects
inherent in any finite difference scheme (with or without artificial dissipation). 
Depending on various parameters such as number of grid points, time step size
 or dissipation factor various effects with different inherent time scales may compete,
and the proper interpretation of results may requires either extremely detailed
and careful parameter studies -- or some analytical modelling in addition to
numerical tests.
As should be expected on theoretical grounds, most codes do require artificial dissipation (e.g. of Kreiss-Oliger type) beyond the linear constant coefficient regime in order to avoid high-frequency instabilities. Particularly clarifying in this respect were Yosef Zlochower's runs with the LazEv BSSN code, and Christiane Lechner's runs with various symmetric hyperbolic codes. As shown in [1],
 it turns out that for second order in space formulations the situation is somewhat more subtle than for first order systems:
while the second derivatives in these systems typically help to damp out high grid frequencies, a mixing of first and second derivatives in the principal part may result in a numerical instability with standard discretizations of certain well posed systems (at least without artificial dissipation).

Since the meeting has taken place, several phone conferences have been organized to further coordinate our work, for information on how to join, news and how to access our data repositories with results see our web site [4].
Finally, let me mention that the steaks {\em are} indeed fabulous in Argentina,
and that they are preferably accompanied by a Malbec from Mendoza.

{\bf References:}

[1] G. Calabrese, I. Hinder and S. Husa, Numerical stability for finite difference approximations of Einstein's equations, \htmladdnormallink{gr-qc/0503056}{http://arxiv.org/abs/gr-qc/0503056}.

[2] B. Reimann, M. Alcubierre, J. A. Gonz\'alez and D. N\'unez, 
 Constraint and gauge shocks in one-dimensional numerical relativity, {\em PRD} {\bf 71},  064021 (2005).

[3] J. Frauendiener and T. Vogel, Algebraic stability analysis of constraint propagation, {\em Class. Quantum Grav.} {\bf 22}, 1769 (2005).

[4] See the project website at 
\htmladdnormallink{\protect {\tt{http://www.appleswithapples.org}}}
{http://www.appleswithapples.org} and the web pages
 of the meeting at
\htmladdnormallink{\protect{\tt{http://www.appleswithapples.org/Meetings/Cordoba2005}}
{http://www.appleswithapples.org/Meetings/Cordoba2005}}.

[5] Here the term ADM refers to the system presented in J. W. York, in Sources of Gravitational Radiation (Cambridge University Press, Cambridge, England, 1979).

[6] M. Alcubierre et al.,  Toward standard testbeds for numerical relativity, {\em Class. Quantum Grav.} {\bf 21}, 589 (2004).

[7] For an overview, new results and further references see Carsten Gundlach, 
Jose M. Martin-Garcia, Hyperbolicity of second-order in space systems of evolution equations, 
\htmladdnormallink{gr-qc/0506037}{http://arxiv.org/abs/gr-qc/0506037}.

\end{document}